\documentclass[amsmath,pra,twocolumn,showpacs]{revtex4}
\usepackage{graphics}

\begin{document}     

\title{Multiorder coherent Raman scattering of a quantum probe field}
\author{Fam Le Kien} 
\altaffiliation{On leave from Department of Physics, University of
Hanoi, Hanoi, Vietnam. Also at Institute of Physics, National Center for 
Natural Sciences and Technology, Hanoi, Vietnam.} 
\affiliation{Department of Applied Physics and Chemistry, 
University of Electro-Communications, Chofu, Tokyo 182-8585, Japan}
\author{Anil K. Patnaik}
\affiliation{Department of Applied Physics and Chemistry, 
University of Electro-Communications, Chofu, Tokyo 182-8585, Japan}
\author{K. Hakuta}
\affiliation{Department of Applied Physics and Chemistry, 
University of Electro-Communications, Chofu, Tokyo 182-8585, Japan}
\date{\today}

\begin{abstract}
We  study the  multiorder  coherent Raman scattering of  
a quantum probe field  in a far-off-resonance medium with a prepared coherence.
Under the conditions of negligible dispersion and limited
bandwidth, we  derive a  Bessel-function solution for the sideband field operators.
We  analytically and numerically calculate various quantum statistical characteristics of
the sideband fields. 
We show that the multiorder coherent Raman
process can replicate the statistical properties of a single-mode quantum probe field into a broad comb of   generated Raman sidebands. 
We also  study the mixing and modulation of photon statistical properties in the case of  two-mode input.   
We  show that   
the prepared Raman coherence and the medium length  can be used as control parameters to switch 
a sideband field from one type of photon statistics to another type, or from a non-squeezed state to a squeezed
state and vice versa.  
\end{abstract}

\maketitle

\section{Introduction}

The  parametric beating of a weak probe field with a prepared Raman
coherence in a far-off-resonance medium has been extensively studied \cite{Liang,Katsuragawa,Nazarkin99,beating}. 
It has  been demonstrated that  multimode laser radiation  \cite{Liang} and 
incoherent fluorescent light \cite{Katsuragawa}  can be  replicated into  Raman sidebands. 
Since a substantial molecular coherence can be produced by the two-color adiabatic Raman pumping method \cite{Modulation,D2,subfem,Kien99}, the quantum conversion efficiency of the parametric beating technique can be maintained high even for weak light with  less than one photon per wave packet \cite{Katsuragawa}. 
To describe the statistical properties of  a weak quantum probe and its first-order Stokes and anti-Stokes sidebands  in the parametric beating process, 
a simplified quantum treatment has recently  been performed \cite{three sidebands}. 
It has been shown  that the statistical  properties of the quantum probe  
can be replicated into the two  sidebands nearest to the input line, in agreement with the experimental observations \cite{Liang,Katsuragawa}.  

However, many experiments have reported the observations of 
ultrabroad Raman spectra with a large number of sidebands \cite{Liang,Katsuragawa,Nazarkin99,Modulation,D2}.
In the experiments with solid hydrogen \cite{Liang,Katsuragawa}, at least two anti-Stokes sidebands and two
Stokes sidebands have been observed.
In the experiment with molecular deuterium \cite{D2}, a large Raman coherence $|\rho_{ab}|\cong 0.33$ and 
about 20 Raman sidebands, covering a wide spectral range from near infrared through vacuum ultraviolet, have been
generated. In  rare-earth doped dielectrics with low Raman frequency  and  long-lived spin coherence, 
a substantial Raman coherence $|\rho_{ab}|\cong 0.25$ and an extremely large number of sidebands  
(about $10^4$)  can also  be generated   \cite{kolesov}.  
Broad combs of Raman sidebands \cite{Liang,Katsuragawa,Nazarkin99,Modulation,D2} have been 
intensively studied because they may  synthesize to  subfemtosecond \cite{subfem,Kien99,Sokolov01,korn02} and subcycle  \cite{HarrisSeries} pulses. 
The generation of broad combs of Raman sidebands has always been examined as a semiclassical problem.
While classical treatments are sufficient for many purposes, 
a quantum treatment is required  when the statistical properties of the radiation fields are important.
On the other hand, broad combs of Raman sidebands with similar nonclassical  properties and
different frequencies may find useful
applications for high-performance optical communication.
Therefore, it is intriguing to examine the quantum aspects of  high-order coherent Raman processes.

In this paper, we extend the treatment of Ref.~\cite{three sidebands} to study various quantum properties of 
multiorder sidebands 
generated by the beating of a quantum probe field with a prepared Raman coherence in a far-off-resonance medium. 
Under the conditions of negligible dispersion and limited
bandwidth, we  derive a  Bessel-function solution for the sideband field operators.
We analytically and numerically calculate various quantum statistical characteristics of the sideband fields
generated from a single-mode quantum input. 
We show that, with increasing the effective medium length or the Raman sideband order, the 
autocorrelation functions, cross-correlation functions, 
photon-number distributions, and squeezing factors undergo oscillations governed by the Bessel functions. 
Meanwhile, the normalized autocorrelation functions and normalized squeezing factors of the single-mode 
probe field are not altered and can be replicated 
into  a broad comb of generated multiorder Raman sidebands. 
We study the mixing and modulation of photon statistical properties in the case of  two-mode input.
We  show that   
the prepared Raman coherence and the medium length can be used as control parameters to switch 
a sideband field from one type of photon statistics to another type, or from a non-squeezed state to a squeezed
state and vice versa. We also discuss two-photon interference in coherent Raman scattering. 
Although the multiorder coherent Raman scattering can produce
a broad comb of sideband fields with different frequencies, 
it behaves in many aspects as a beam splitter 
\cite{coupler,Mandel and Scully book,beam splitter,Hong,applications,Knight}. 
Therefore, in this paper, we also make comparison of this conventional device
with our system as and when it is possible.
 
Before we proceed, we note that, in related problems, 
the generation of correlated photons using the $\chi^{(2)}$ and $\chi^{(3)}$ parametric processes 
has  been studied \cite{Mandel and Scully book,coupler,Wang}. 
The correlations between the Stokes and anti-Stokes sidebands and 
the possibility of  transferring a quantum state of light from one carrier
frequency  to another carrier  frequency (multiplexing) have been discussed for resonant systems \cite{Scully}.

The paper is organized as follows. In Sec.\ \ref{sec:model}, we describe the model and
present the basic equations. 
In Sec.\ \ref{sec:single-mode}, we study various quantum characteristics of the sideband fields generated
from a single-mode quantum input. 
In Sec.\ \ref{sec:two-mode}, we discuss the quantum properties of the sideband fields  generated
from a two-mode quantum input. 
Finally, we present the conclusions in Sec.\ \ref{sec:summary}.

\section{Model}
\label{sec:model}

\begin{figure}
\begin{center}
  \includegraphics{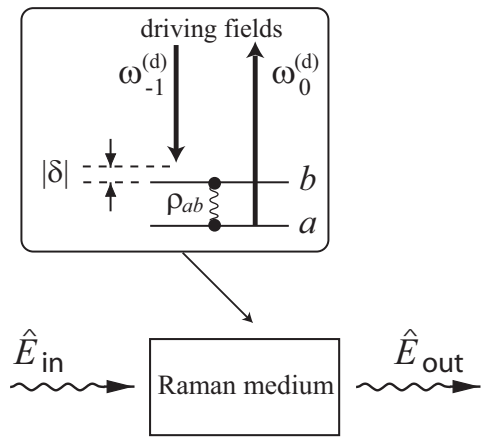}
 \end{center}
\caption{Principle of the technique: 
Two classical laser fields drive 
a Raman transition of molecules in a  far-off-resonance medium. 
The beating of a weak quantum probe field with the prepared Raman coherence produces  new sideband fields.}
\label{fig1}
\end{figure}

We consider a far-off-resonance  Raman medium shown schematically in Fig.~\ref{fig1}. 
Level $a$ with energy $\omega_a$ is coupled to  
level  $b$ with energy $\omega_b$ by a Raman transition via intermediate levels that are not shown in the figure. 
We send a pair of long, strong, classical  laser fields, with carrier frequencies 
$\omega_{-1}^{(d)}$ and $\omega_0^{(d)}$,
and a short, weak, quantum probe field $\hat E_{\mathrm{in}}$, with one or several 
carrier frequencies, through the  Raman medium, along the $z$ direction. The timing and alignment of these fields are such  that  they  
substantially overlap with each other 
during the interaction process. The driving laser fields are tuned close to 
the Raman transition $a \leftrightarrow b$, with a small finite two-photon detuning $\delta$,  
but are far detuned from the upper electronic states $j$ of the molecules. 
We assume that all the frequency components of the input probe field   are separated 
by integer multiples of the Raman modulation frequency $\omega_m=\omega_b-\omega_a-\delta$.  
The driving fields adiabatically 
produce a Raman coherence $\rho_{ab}$ \cite{subfem,Kien99}. 
When the probe field propagates through the medium, it beats with  the  prepared
Raman coherence.
Since the probe field is weak and short compared to the driving fields, 
the medium state and the driving fields do not change substantially during this step. 
The beating of the probe field with the prepared Raman coherence leads to the generation of  new  sidebands in the total  output field $\hat E_{\mathrm{out}}$. 
The frequencies of the  sideband fields $\hat E_q$ are given by 
$\omega_q=\omega_0+q\omega_m$, where 
$q$ is integer and $\omega_0$ is a carrier frequency of the input probe field. 
The range of $q$ should be appropriate so that $\omega_q$ is positive.
The probe field is taken to be not too short so that the Fourier-transformation limited broadening is negligible. 
We assume that the prepared Raman coherence $\rho_{ab}$ is substantial  so that the spontaneous Raman process
is negligible compared to the stimulated and parametric processes. Consequently, the quantum noise
can be neglected. 
Unlike Ref.~\cite{three sidebands}, our model does not
require any restriction on the magnitude of the coherence as all Raman sidebands are included. 
When we take the propagation equation for the classical Raman sidebands \cite{subfem,Kien99}
and replace the  field amplitudes   by the quantum operators, we obtain 
\begin{equation}
\frac{\partial \hat E_q}{\partial  z }+\frac{\partial \hat E_q}{c\partial  t }
=i\beta_q
(u_q\hat E_q +d_{q-1}\rho_{ba}\hat E
_{q-1} +d_q\rho_{ab}\hat E_{q+1}).
\label{1}
\end{equation}
Here, $u_q$ and $d_q$  are the dispersion and coupling constants, respectively. We have denoted 
$ \beta_q= {\cal N} \hbar \omega_q/\epsilon_0 c$,  where ${\cal N}$ is the molecular number density.

We take all  the sidebands to be sufficiently far from  resonance  that
the dispersion of the medium is negligible. In this case, we have  
$u_q=u_0$ and $d_q=d_0$. 
We write $\rho _{ab} = \rho_0 \exp[i(\phi_0-\beta_m u_0 z )]$, 
where $\rho_0\equiv|\rho_{ab}|$ and $ \beta_m= {\cal N} \hbar \omega_m/\epsilon_0 c$, and assume that $\rho_0$ and $\phi_0$ are constant in time and space.
We change the variables by $\hat E_q=\hat {\cal E}_q \exp[i (\beta_q u_0  z-q\phi_0) ]$.
Using  photon operators,  we can write
$\hat{\cal E}_q(z,t)=(2\hbar\omega_q/\epsilon_0LA)^{1/2}\sum_K  \hat b_q(K,t) e^{iK(z-ct)}$.
Here, $L$ is the quantization length taken to be equal to the medium length, 
$A$ is the quantization transverse area taken to be equal to the beam area, 
$K$ is a Bloch wave vector, and $\hat b_q(K,t)$ and $\hat b_q^\dagger(K,t)$ are the annihilation and creation operators for photons in the spectral mode $q$ and the spatial mode $K$.
For simplicity, we restrict our discussion to  the case where each sideband field contains 
only a single spatial mode  (with, e.g., $K=0$). 
Then, Eq.~(\ref{1}) yields  
\begin{equation}
\frac{\partial \hat b_q}{\partial  t }
=i(g_q \hat b_{q-1} +g_{q+1} \hat b_{q+1}),
\label{4}
\end{equation}
where $g_q=({\cal N} \hbar/\epsilon_0 )\sqrt{\omega_q\omega_{q-1}} \, d_0  \rho_0$. 
For the  medium length $L$, the evolution time is $t=L/c$. 
It follows from Eq.~(\ref{4}) that  the  total photon number is conserved in time.
Note that Eq.~(\ref{4}) 
represents the Heisenberg equation for the fields that are coupled to each other by the effective interaction Hamiltonian
\begin{equation}
\hat H=
-\hbar \sum_q g_{q+1} (\hat b_q \hat b_{q+1}^\dagger+\hat b_{q+1} \hat b_q^\dagger).
\label{8b}
\end{equation}
The interaction between  the sideband fields  via the prepared Raman coherence  is analogous to the  interaction between the transmitted and
reflected fields from a conventional beam splitter \cite{Mandel and Scully book}. 
However,  the two mechanisms are very different in physical nature. 
The most important
difference between them is that the two fields from the conventional beam splitter have the same
frequency while the sideband fields in the  Raman scheme 
have different frequencies. In addition, the model Hamiltonian (\ref{8b}) involves an infinitely large number of Raman sidebands, separated by integer multiples of the Raman modulation frequency $\omega_m$. 
Despite these differences, the model (\ref{8b}) can be called the multiorder Raman beam splitter.
The parameters  $g_qt=g_qL/c$ determine the transmission and scattering coefficients for
the fields at the Raman beam splitter.

We assume that the bandwidth of the generated Raman spectrum is small compared to the 
characteristic probe frequency $\omega_0$.
In this case, the $q$-dependence of the coupling parameters 
$g_q$  can be neglected, that is, we have
$g_q=g_0={\cal N} \hbar \omega_0d_0\rho_0/\epsilon_0$.
With this assumption, we find the following solution to Eq.~(\ref{4}):
\begin{equation} 
\hat b_q(t)=\sum_{q'} e^{i(q-q')\pi/2} J_{q-q'}(2g_0t)\hat b_{q'}(0).
\label{8}
\end{equation}
Here, $J_k$ is the $k$th-order Bessel function. 
The expression (\ref{8}) for the output field operators  is a generalization of the Bessel-function solution 
obtained earlier for the classical fields  \cite{subfem,Kien99}. The  number of generated Raman sidebands is characterized by the effective interaction time $2g_0t$ or, equivalently,
the effective medium length $\kappa L$, where   
\begin{equation}
\kappa=\frac{2g_0}{c}=\frac{2\hbar }{\epsilon_0c}{\cal N}\omega_0d_0\rho_0. 
\end{equation}
The coefficient $\kappa$ characterizes the strength of the parametric coupling and is proportional to the prepared Raman coherence $\rho_0$, that is, to the intensities of the driving laser fields. 
The Bessel functions $J_k(\kappa L)$ are the transmission ($k=0$) and scattering ($k\not=0$) coefficients for
the Raman sidebands, similar to the transmission and  reflection 
coefficients of a conventional beam splitter. 
The assumption of limited bandwidth requires $\kappa L\ll\omega_0/\omega_m$, that is,
$(2\hbar/\epsilon_0c){\cal N}\omega_md_0\rho_0L\ll 1$ \cite{subfem,Kien99}.
In what follows we use the  expression (\ref{8}) to calculate  various quantum statistical characteristics of the   sideband  fields, namely, the autocorrelation functions, the two-mode  cross-correlation functions,  the squeezing  
factors, and the relation between the $P$ representations of the output and input states.

\section{Single-mode quantum input}
\label{sec:single-mode}

In this section, we consider the case where the input probe field has a single carrier frequency $\omega_0$.
In other words,
we assume that the sideband $q=0$ is initially prepared in a quantum state $\hat\rho_{\mathrm{in}}^{(0)}$ and the other sidebands are initially in the vacuum state. The density matrix of the initial  state of the  fields is given by 
\begin{equation}
\hat\rho_{\mathrm{in}}=\hat\rho_{\mathrm{in}}^{(0)}\otimes\prod_{q\not=0} (|0\rangle\langle 0|)_q.
\label{8a}
\end{equation}

\subsection{Autocorrelation functions}

We study the autocorrelations of photons in the generated Raman sidebands.
We use Eq.~(\ref{8}) and apply the initial density matrix (\ref{8a}) to calculate the normally ordered  moments 
$\langle\hat b_q^{\dagger n}\hat b_q^n \rangle$ of the photon-number operators 
$\hat n_q=\hat b_q^\dagger \hat b_q$. The result is
\begin{equation}
\langle\hat b_q^{\dagger n}\hat b_q^n \rangle=J_q^{2n}(\kappa L)
\langle\hat b_0^{\dagger n}(0)\hat b_0^n(0)\rangle.
\label{9}
\end{equation}
In particular, the mean photon numbers of the sidebands are given by
\begin{equation}
\langle\hat n_q \rangle=J_q^2(\kappa L)\langle\hat n_{\mathrm{in}}\rangle.
\label{9a}
\end{equation}
Here, $\hat n_{\mathrm{in}}=\hat b_0^\dagger(0)\hat b_0(0) $ is the photon-number operator for 
the input field.
The $n$th-order autocorrelation functions of the sidebands are defined by 
$\Gamma_q^{(n)}=\langle\hat b_q^{\dagger n}\hat b_q^n \rangle-\langle\hat b_q^\dagger\hat b_q \rangle^n$.   
From Eqs.~(\ref{9}) and (\ref{9a}), we find  
\begin{equation}
\Gamma_q^{(n)}=J_q^{2n}(\kappa L)\Gamma_{\mathrm{in}}^{(n)},  
\label{9b}
\end{equation}
where $\Gamma_{\mathrm{in}}^{(n)}=\langle\hat b_0^{\dagger n}(0)\hat b_0^n (0)\rangle
-\langle\hat b_0^\dagger(0)\hat b_0 (0)\rangle^n$ is the $n$th-order autocorrelation function of the input field.   

\begin{figure}
\begin{center}
\includegraphics{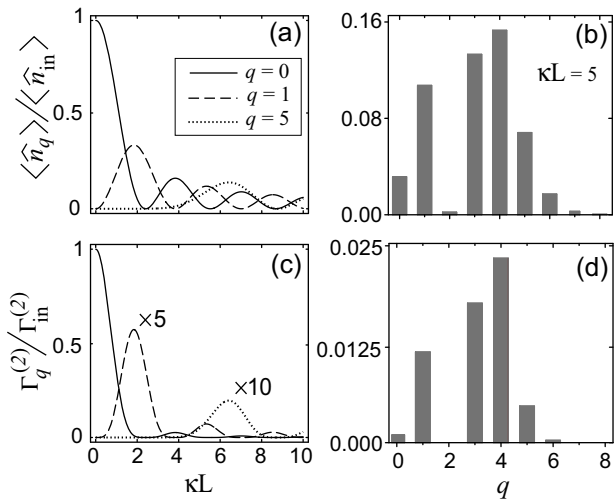}
\end{center}
\caption{
Mean photon number $\langle \hat n_q\rangle$ (first row) and second-order autocorrelation function $\Gamma_q^{(2)}$ (second row), both scaled to their initial values for the probe field,
as functions of the effective medium length $\kappa L$ (first column) and the sideband order $q$ (second column).
In (a) and (c), the sideband order is $0$ (solid line), $1$ (dashed line), and $5$ (dotted line).
In (b) and (d), the effective medium length is $\kappa L = 5$.
In (c), we have amplified $\Gamma_1^{(2)}/\Gamma_{\mathrm{in}}^{(2)}$ (dashed line) and $\Gamma_5^{(2)}/\Gamma_{\mathrm{in}}^{(2)}$ (dotted line) by 5 and 10 times, respectively. 
In (b) and (d), the  negative side of the $q$ axis is not shown  because the functions plotted are symmetric in $q$.} 
\label{Fig2}
\end{figure}

Equations (\ref{9a}) and (\ref{9b}) indicate that, 
when we increase the effective medium length $\kappa L$ or the sideband order $q$, 
the mean photon number $\langle \hat n_q\rangle$ and the autocorrelation function $\Gamma_q^{(n)}$ 
undergo oscillations as described by even powers of the Bessel function $J_q(\kappa L)$.
Such  oscillatory behavior is illustrated in Fig.~\ref{Fig2}.
When the  sideband order $q$ is higher,  the  onset of $\langle \hat n_q\rangle$ occurs later [see Fig.~\ref{Fig2}(a)]
and, hence, so does  the onset of $\Gamma_q^{(n)}$ [see Fig.~\ref{Fig2}(c)].
For a fixed $q$, both $\langle \hat n_q\rangle$ and $\Gamma_q^{(n)}$
reach their largest values at the same optimal medium length $L_q=x_q/\kappa$, where $x_q$ is the position of 
the first peak of $J_q(x)$. 
The higher the sideband order $q$, the larger is the optimal length $L_q$ and the smaller are the  maximal output values of $\langle \hat n_q\rangle$ and  $\Gamma_q^{(n)}$ [see Figs.~\ref{Fig2}(a) and \ref{Fig2}(c)]. 
Figures \ref{Fig2}(b) and \ref{Fig2}(d)  show that  $\langle \hat n_q\rangle$ and $\Gamma_q^{(n)}$ 
are substantially different from zero only in the region where $|q|$ is not too large compared to 
$\kappa L$. For a given $\kappa L$, both 
$\langle \hat n_q\rangle$ and $\Gamma_q^{(n)}$ achieve their maximal values at $q\approx \kappa L$.

The normalized $n$th-order autocorrelation functions of the sidebands are defined by 
$g_q^{(n)}=\langle\hat b_q^{\dagger n}\hat b_q^n \rangle /\langle\hat b_q^\dagger\hat b_q \rangle^n$.
These functions characterize the overall statistical  properties, such as
sub-Poissonian, Poissonian, or super-Poissonian photon statistics, regardless of the mean photon number.
Unlike the mean photon number $\langle \hat n_q\rangle$ and  
the autocorrelation function $\Gamma_q^{(n)}$, 
the normalized autocorrelation function $g_q^{(n)}$
does not oscillate when we change  the effective medium length $\kappa L$ or the sideband order $q$. 
Indeed, with the help of Eq.~(\ref{9}), we find 
\begin{equation}
g_q^{(n)}=g_{\mathrm{in}}^{(n)},
\label{10}
\end{equation}
where $g_{\mathrm{in}}^{(n)}=
\langle\hat b_0^{\dagger n}(0)\hat b_0^n(0) \rangle/\langle\hat b_0^\dagger(0)\hat b_0(0) \rangle^n$.

Equation (\ref{10}) indicates that the generated sideband fields and the probe field have  the same normalized autocorrelation functions, which are independent of the evolution time and are solely determined by the statistical properties of the input field. 
In other words, the normalized autocorrelation functions of the probe field do not change during 
the parametric beating
process and are precisely replicated into the comb of generated sidebands. 
Such a replication of the normalized autocorrelation characteristics can be called autocorrelation multiplexing.
In particular, if the photon statistics of the input field is sub-Poissonian, Poissonian, or
super-Poissonian, the photon statistics of each of  the sideband fields will also be sub-Poissonian, Poissonian,
or super-Poissonian, respectively.
This result is in agreement with the experiments on replication
of multimode laser radiation \cite{Liang} and broadband incoherent light \cite{Katsuragawa}.
The ability of the Raman medium to multiplex the  autocorrelation characteristics
is similar to the ability of a conventional beam splitter \cite{Mandel and Scully book}.

It is not surprising that the normalized autocorrelation functions of the probe field are replicated into the sidebands in the parametric beating process.
Such a replication is possible because  
the medium is far off resonance and the quantum probe field is weak
compared to the driving fields. 
Under these two conditions,  
the photon annihilation operators are linearly transformed 
as described by the linear differential  equation (\ref{4}). 
This equation shows that the photon annihilation operators
are not mixed up with the creation operators, 
and therefore the evolution of the annihilation operators is linear with respect to  
the initial annihilation operators.

We should, however, emphasize here that the replication of all of the normalized autocorrelation functions $g_{\mathrm{in}}^{(n)}$ (for all orders $n$)  of the input field 
does not mean the replication of the quantum state $\hat\rho_{\mathrm{in}}^{(0)}$. 
In fact, the oscillations of the mean photon number
$\langle \hat n_q\rangle$ and the autocorrelation function $\Gamma_q^{(n)}$ indicate that
the photon-number distributions and consequently the quantum states of the sidebands evolve in a rather complicated way and are quite different from those for the input field.
A separable state at the input can produce an entangled state \cite{three sidebands,entang}.
In addition, 
cross-correlations between the sidebands can be generated from  initially uncorrelated fields,
and a Fock state at the input
does not produce  sideband fields  in isolated Fock states [see below].

\subsection{Cross-correlation functions}

\begin{figure}
\begin{center}
\includegraphics{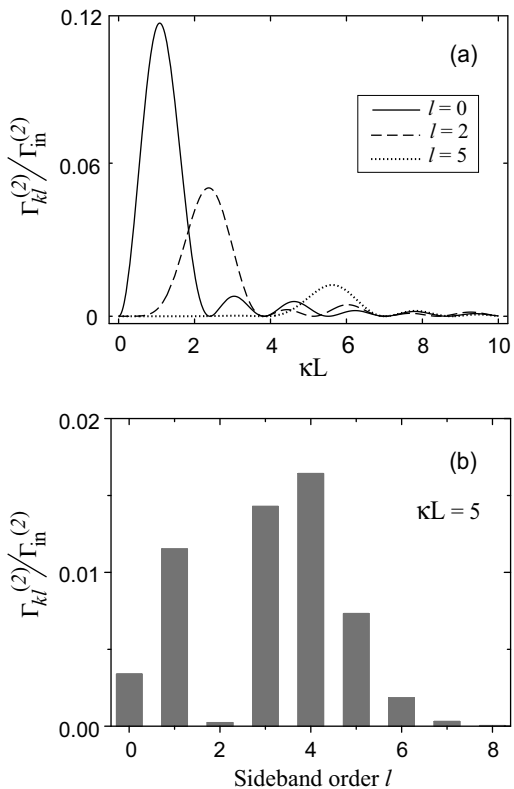}
\end{center}
\caption{
Cross-correlation function $\Gamma_{kl}^{(2)}$, scaled with the second-order autocorrelation function of the input field, as a function of (a) the effective medium length $\kappa L$  and (b) the sideband order $l$.
Here, the sideband order $k$ is fixed  to  $1$. 
In (a), the sideband order $l$ is $0$ (solid line), $2$ (dashed line), and $5$ (dotted line).
In  (b),  the effective medium length is $\kappa L = 5$.
In (b), the negative side of the $l$ axis is not shown  
because the function plotted is symmetric with respect to the sideband orders.} 
\label{Fig3}
\end{figure}

We study the correlations between the generated Raman sidebands.
For two different sidebands $k$ and $l$ ($k\not=l$), we have 
\begin{equation}
\langle\hat n_k\hat n_l\rangle=J_k^2(\kappa L)J_l^2(\kappa L)\langle\hat n_{\mathrm{in}}(\hat n_{\mathrm{in}}-1)\rangle.
\label{11b}
\end{equation}
The  cross-correlation function for the two sidebands is defined by $\Gamma_{kl}^{(2)}=\langle\hat n_k\hat n_l\rangle-\langle\hat n_k\rangle\langle\hat n_l\rangle$. Using Eqs.~(\ref{9a}) and (\ref{11b}), we find  
\begin{equation}
\Gamma_{kl}^{(2)}=J_k^2(\kappa L)J_l^2(\kappa L)\Gamma_{\mathrm{in}}^{(2)}.
\label{11a}
\end{equation}
When we extend $\Gamma_{kl}^{(2)}$ for $k=l$, we have $\Gamma_{kk}^{(2)}= \Gamma_k^{(2)}$.
According to Eq.~(\ref{11a}), the  cross-correlation function $\Gamma_{kl}^{(2)}$ oscillates when we change the effective medium length $\kappa L$ or the sideband orders $k$ and $l$.  
Such  oscillatory behavior is illustrated in Fig.~\ref{Fig3}.

The normalized cross-correlation function is defined by
$g_{kl}^{(2)}=\langle\hat n_k\hat n_l\rangle/(\langle\hat n_k\rangle\langle\hat n_l\rangle)$. 
Unlike the function $\Gamma_{kl}^{(2)}$, the normalized  function
$g_{kl}^{(2)}$ does not oscillate.
Indeed, we find the relation
\begin{equation}
g_{kl}^{(2)}=g_{q}^{(2)}=g_{\mathrm{in}}^{(2)}.
\label{11}
\end{equation}
As seen from the above relation, the normalized  cross-correlation functions $g_{kl}^{(2)}$  for all possible sideband pairs
$(k,l)$   are equal to each other, to the normalized second-order 
autocorrelation function $g_{q}^{(2)}$ for each sideband $q$, 
and  to the normalized second-order autocorrelation function 
$g_{\mathrm{in}}^{(2)}$ of the input field. 
When $g_{\mathrm{in}}^{(2)}\not=1$, that is, when the photon statistics of the input field is non-Poissonian, we obtain $g_{kl}^{(2)}\not=1$, a signature of cross-correlations between the sidebands.
Such correlations are generated although the sidebands are initially not correlated. 
In particular, if  the input field has a sub-Poissonian photon
statistics ($g_{\mathrm{in}}^{(2)}<1$), anti-correlations between the sidebands ($g_{kl}^{(2)}<1$) will be generated. 
Note that the conventional beam splitters also have a similar property \cite{Mandel and Scully book}.
The generation of cross-correlations between the sidebands indicates that
the quantum states of the generated sidebands are different from that of the input field.
We emphasize that the  anti-correlation generation  cannot be explained 
by the classical statistics of the fields with positive $P$ functions  although the sideband dynamics  is linear with respect to the field variables.

\subsection{Photon-number distributions}
\label{distribution}

To get deeper insight into the quantum properties of the generated Raman sidebands, 
we derive the photon-number distributions  for the output
fields. The joint photon-number distribution for the output fields is defined by $P_{\Sigma}(\{n_q\})=\langle\{n_q\}|\hat\rho_{\mathrm{out}}|\{n_q\}\rangle$,
where $\hat\rho_{\mathrm{out}}=\hat U(L/c)\hat\rho_{\mathrm{in}}\hat U^\dagger(L/c)$ is the density matrix of the output state. 
Here, $\hat U(L/c)=\exp(-i\hat HL/\hbar c)$ is the evolution operator.
Using Eqs.~(\ref{8}) and (\ref{8a}), we find
\begin{equation}
P_{\Sigma}(\{n_q\})=  p_{\mathrm{in}}(N) \frac{N!}{\prod_q n_q!} \prod_q  J_{q}^{2n_q}(\kappa L),
\label{10a}
\end{equation} 
where $p_{\mathrm{in}}(n)= \mbox{}_0\langle n|\hat\rho_{\mathrm{in}}^{(0)}|n\rangle_0$ is
the  photon-number distribution  of the input field, and $N=\sum_q n_q$ is the total photon number.
From Eq.~(\ref{10a}), the marginal photon-number distribution $p_q(n)$ for the sideband $q$
is obtained as 
\begin{equation}
p_q(n)=\frac{J_{q}^{2n}(\kappa L)}{n!} \sum_{k=0}^\infty
\frac{(n+k)!}{k!}
[1-J_{q}^2(\kappa L)]^k 
p_{\mathrm{in}}(n+k).
\label{10b}
\end{equation}
Clearly,  $p_q(n)$ is in general different from $p_{\mathrm{in}}(n)$. Thus, despite the replication of the normalized autocorrelation functions, the photon-number distribution 
of the probe field is not  replicated into the sidebands. 

We examine several particular cases.
First, we consider the case where the probe field is initially prepared in a coherent state $|\alpha\rangle_0$.
This state is characterized by a Poisson distribution $p_{\mathrm{in}}(n)=e^{-\bar{N}} \bar{N}^n/n!$ for the photon number, where $\bar{N}=|\alpha|^2$. 
In this case, Eqs.~(\ref{10a}) and (\ref{10b}) yield $P_{\Sigma}(\{n_q\})=\prod_q p_q(n_q)$ and 
\begin{equation} 
p_q(n)=e^{-\bar{n}_q} \frac{ \bar{n}_q^n}{n!},
\end{equation} 
respectively, where $\bar{n}_q=\bar{N}J_q^2(\kappa L)$. 
Thus, the generated sidebands are not correlated, and the marginal photon-number distributions for the individual sidebands
remain Poisson distributions during the evolution  process. 

Second, we consider the case where the probe field is initially in a Fock state $|N\rangle_0$.
In this case, we have $p_{\mathrm{in}}(n)=\delta_{n,N}$. 
Therefore, Eq.~(\ref{10b}) yields 
\begin{equation}
p_q(n)=J_{q}^{2n}(\kappa L) 
[1-J_{q}^2(\kappa L)]^{N-n} 
\frac{N!}{n!(N-n)!}
\end{equation} 
for $n\leq N$,
and $p_q(n)=0$ for $n > N$. 
Meanwhile, Eq.~(\ref{10a}) yields 
$P_{\Sigma}(\{n_q\})\not=\prod_q p_q(n_q)$, that is, the joint photon-number distribution  is not a product of the marginal 
photon-number distributions for the individual sidebands. Thus, 
the sidebands are correlated. They are not generated in isolated Fock states.    
   
Finally, we consider the case where the probe field is initially in a thermal state, which is characterized by
a Boltzmann photon-number distribution 
$p_{\mathrm{in}}(n)=\bar{N}^n/(\bar{N}+1)^{n+1}$. In this case, 
Eq.~(\ref{10b}) yields 
\begin{equation}
p_q(n)=\frac{\bar{n}_q^n}{(\bar{n}_q+1)^{n+1}},
\end{equation} 
where $\bar{n}_q=\bar{N}J_q^2(\kappa L)$. 
Thus, the marginal photon-number distributions for the individual sidebands
remain Boltzmann distributions during the evolution process. 
However, according to Eq.~(\ref{10a}), we have
$P_{\Sigma}(\{n_q\})\not=\prod_q p_q(n_q)$, a signature of correlations between the generated sidebands.

\subsection{Squeezing}
\label{squeezing}

We examine  the  squeezing  of the  field quadratures.
A field quadrature of the $q$th mode is defined by 
$\hat X_q=\hat b_q^\dagger e^{i\varphi}+\hat b_qe^{-i\varphi}$.
We say that the $q$th  mode is in a squeezed state if there exists such a phase $\varphi$ that 
$\langle(\Delta \hat X_q)^2\rangle < 1$ or, equivalently, $S_q<0$, where 
$S_q=\langle(\Delta \hat X_q)^2\rangle - 1$. 
The squeezing degree is measured by the quantity $-S_q$.
Note that the relation between the squeezing factor $S_q$ and the conventional squeezing parameter 
$r_q$  is $S_q=e^{-2r_q}-1$.
In terms of the photon operators, we have 
\begin{equation}
S_q=2[\langle \hat b_q^\dagger\hat b_q\rangle-\langle\hat b_q^\dagger\rangle\langle\hat b_q\rangle]
+[(\langle \hat b_q^2\rangle-\langle \hat b_q\rangle^2) 
e^{-2i\varphi}+\mathrm{c.c.}].
\label{12c}
\end{equation}
Using Eqs.~(\ref{8}) and (\ref{8a}), we find
\begin{equation} 
\langle\hat b_q\rangle= e^{iq\pi/2} J_{q}(\kappa L)\langle\hat b_0(0)\rangle
\label{12a}
\end{equation}
and
\begin{equation} 
\langle\hat b_q^2\rangle= e^{iq\pi} J_q^2(\kappa L)\langle\hat b_0^2(0)\rangle.
\label{12b}
\end{equation}
When we insert Eqs.~(\ref{9a}), (\ref{12a}), and (\ref{12b}) into Eq.~(\ref{12c}), we obtain
\begin{equation}
S_q(\varphi+q\pi/2)=J_q^2(\kappa L) S_{\mathrm{in}}(\varphi).
\label{12}
\end{equation}
Here,  $S_{\mathrm{in}}(\varphi)$ denotes the squeezing factor for the $\varphi$-quadrature of the input field. 
Equation (\ref{12}) shows that, if $S_{\mathrm{in}}(\varphi)<0$, then $S_q(\varphi+q\pi/2)<0$. Thus, if the input field is in a squeezed state,  then the generated sidebands are also in squeezed states. In other words, the squeezing of the input 
field is transferred to the comb of  generated sidebands during the parametric beating process.
The squeezing factor $S_q(\varphi+q\pi/2)$ of the sideband $q$
is reduced from the input squeezing factor $S_{\mathrm{in}}(\varphi)$  by the factor $J_q^2(\kappa L)$. 
Unlike the case of  linear directional couplers and beam splitters  \cite{coupler}, 
the squeezing degree of the probe field cannot be completely transferred to the Raman sidebands. 
This difference is due to the fact that the linear directional coupler and the beam splitter involve only two output modes while the multiorder coherent Raman process involves 
many more output modes.
Note that the phase of the squeezed quadrature of the sideband $q$  changes by $q\pi/2$.
This means that the squeezed quadrature of a generated even-order (odd-order) Raman sideband  
is parallel (orthogonal) to that of the input field.

We introduce the normalized squeezing factor $s_q=S_q/\langle \hat n_q\rangle$. We find
the relation  
\begin{equation}
s_q(\varphi+q\pi/2)=s_{\mathrm{in}}(\varphi),
\label{12d}
\end{equation}
where $s_{\mathrm{in}}(\varphi)=S_{\mathrm{in}}(\varphi)/\langle\hat n_{\mathrm{in}}\rangle$ is the normalized squeezing factor for the input field.
Equation (\ref{12d}) implies that, besides a shift of the quadrature phase angle, 
the normalized squeezing factor for the input field
is replicated into the comb of generated sidebands.
This result can be used to  convert 
squeezing to a new frequency, i.e., to perform squeezing multiplexing.
The relation $S_q(\varphi+q\pi/2)/S_{\mathrm{in}}(\varphi)=\langle\hat n_q \rangle/\langle\hat n_{\mathrm{in}}\rangle$ indicates that the $\kappa L$- and $q$-dependences of the squeezing factor $S_q(\varphi+q\pi/2)$  are similar to those
of the mean photon number $\langle\hat n_q \rangle$ [see Figs.~\ref{Fig2}(a) and \ref{Fig2}(b)].   
Note that, when the input field is in a coherent state, the sideband fields have no squeezing. This property is similar to the case of four-wave mixing but is unlike
the case of degenerate parametric down-conversion, where perfect squeezing can  in principle be obtained. 
Since squeezed states are nonclassical states, the ability of the Raman medium to multiplex squeezing
from a probe field to its sidebands is a quantum property that 
cannot be described by the classical statistics of the fields with positive $P$ functions.

\subsection{Quantum states of the output fields}
\label{coherent multiplexing}

We  calculate the quantum state of the output fields for several  cases. First, we  consider
the case where the input sideband $0$ is initially in a coherent state $|\alpha\rangle_0$.
The  state of the fields at the input is 
\begin{equation}
|\Psi_{\mathrm{in}}\rangle=|\alpha\rangle_0\prod_{q\not=0}|0\rangle_q =e^{-|\alpha|^2/2}e^{\alpha \hat b_0^\dagger(0)}|0\rangle.
\label{13}
\end{equation}
The state of the fields at the output is given by 
$|\Psi_{\mathrm{out}}\rangle=\hat U(L/c) |\Psi_{\mathrm{in}}\rangle$.
Since $\hat U(L/c)|0\rangle=|0\rangle$, we have
$|\Psi_{\mathrm{out}}\rangle=e^{-|\alpha|^2/2}e^{\alpha \hat b_0^\dagger(-L/c)}|0\rangle$.
Using Eq.~(\ref{8}), we find 
\begin{equation}
|\Psi_{\mathrm{out}}\rangle=|\{\alpha_q(L/c)\}\rangle\equiv\prod_q |\alpha_q(L/c)\rangle_q. 
\label{14}
\end{equation}
Here, $|\alpha_q(L/c)\rangle_q$ is a coherent state of the $q$th mode, with the amplitude   
\begin{equation}
\alpha_q(L/c)=\alpha J_{q}(\kappa L) e^{iq\pi/2}. 
\label{15}
\end{equation}
Thus, a probe field in a  coherent state 
can produce  sideband fields that are also in coherent states. 
Such a process can be called   coherent-state multiplexing.
This  property of the Raman medium is  similar to the case of  conventional beam splitters \cite{Mandel and Scully book}.

Second, we consider the case where the input sideband $0$ is initially prepared in a Fock state $|N\rangle_0$. 
The  state of the fields at the input is written as 
\begin{equation}
|\Psi_{\mathrm{in}}\rangle=|N\rangle_0\prod_{q\not=0}|0\rangle_q =
\frac{1}{\sqrt{N!}}\hat b_0^{\dagger N}(0)|0\rangle.
\label{15c}
\end{equation}
The output state of the fields is given by $|\Psi_{\mathrm{out}}\rangle=
(N!)^{-1/2}\hat b_0^{\dagger N}(-L/c)|0\rangle$.
With the help of Eq.~(\ref{8}), we find 
\begin{equation}
|\Psi_{\mathrm{out}}\rangle=\sum_{\{n_q\}} C_{\{n_q\}}^{(N)}|\{n_q\}\rangle. 
\label{15a}
\end{equation}
Here,  
\begin{equation}
C_{\{n_q\}}^{(N)}=\sqrt{\frac{N!}{\prod_q n_q!}}\prod_q e^{iqn_q\pi/2}J_q^{n_q}(\kappa L)
\label{15b}
\end{equation}
for $\sum_q n_q=N$, and $C_{\{n_q\}}^{(N)}=0$ for $\sum_q n_q\not=N$. When $N\not=0$
and $\kappa L\not=0$, the output
state (\ref{15a}) is, in general, a multipartite inseparable (entangled) state. 

In a particular case where the input state of the probe field is a single-photon state, i.e., $N=1$,
Eqs.~(\ref{15a}) and (\ref{15b}) yield 
\begin{equation}
|\Psi_{\mathrm{out}}\rangle=\sum_{q} e^{iq\pi/2}J_q(\kappa L)|1_q\rangle.
\end{equation}
Here, $|1_q\rangle$ is the quantum state  of a single photon  in the sideband $q$ with no photons in the other sidebands. In this case, the entanglement between two different  sidebands $k$ and $l$ at the output can be measured by the bipartite concurrence  $C_{kl}=2|J_k(\kappa L)J_l(\kappa L)|$, see \cite{entang}. 

Finally, we consider the case
where the input sideband $0$ is initially in an incoherent mixed state 
\begin{equation}
\hat\rho_{\mathrm{in}}^{(0)}=\sum_n p_{\mathrm{in}}(n)(|n\rangle\langle n|)_0.
\end{equation}
With the help of Eq.~(\ref{15a}), the density matrix of the output state of the fields  is found to be
\begin{equation}
\hat\rho_{\mathrm{out}}=
\sum_{N,\{n_q\},\{n'_q\}} p_{\mathrm{in}}(N)
C_{\{n_q\}}^{(N)}C_{\{n'_q\}}^{(N)*}|\{n_q\}\rangle\langle\{n'_q\}|.
\label{15d}
\end{equation}
To obtain the reduced density matrix $\hat\rho_{\mathrm{out}}^{(q)}$ 
for an arbitrary sideband $q$,  we trace the total density matrix (\ref{15d})  over all sidebands except
for the sideband $q$. Then, we find
\begin{equation}
\hat\rho_{\mathrm{out}}^{(q)}=\sum_{n} p_q(n) (|n\rangle\langle n|)_q,
\end{equation}
where the marginal photon-number distribution $p_q(n)$ is given by Eq.~(\ref{10b}).
As seen, the reduced state of each sideband is also an incoherent superposition of Fock states.
However, if the input  state  is different from the vacuum state, then, for $\kappa L\not=0$,
we have $\hat\rho_{\mathrm{out}}\not=\prod_q \hat\rho_{\mathrm{out}}^{(q)}$,
a signature of correlations between the generated sidebands. 
Moreover, the total density matrix (\ref{15d}) of the output fields contains nonzero off-diagonal matrix elements in the Fock-state basis. 

In a particular case where the initial state of the probe field  is a thermal state,
i.e., $\hat\rho_{\mathrm{in}}^{(0)}=\sum_n [\bar{N}^n/(\bar{N}+1)^{n+1}](|n\rangle\langle n|)_0$,
the reduced state of each generated sideband is also a thermal state, namely,  
\begin{equation}
\hat\rho_{\mathrm{out}}^{(q)}=\sum_n \frac{\bar{n}_q^n}{(\bar{n}_q+1)^{n+1}}(|n\rangle\langle n|)_q.
\end{equation}
Here, $\bar{n}_q=\bar{N}J_q^2(\kappa L)$ is the mean photon number for the sideband $q$. 
The reduced thermal states of the generated sidebands are, however, not isolated from each other.

\section{Two-mode quantum input}
\label{sec:two-mode}

A far-off-resonance medium with a substantial Raman coherence, 
prepared by two strong driving fields,  can efficiently mix and modulate 
the quantum statistical properties of the sideband fields. 
To understand this mechanism, we study the case 
where the input probe field has two carrier frequencies, 
$\omega_0$ and $\omega_\nu=\omega_0+\nu\omega_m$, separated by an  integer multiple 
$\nu$ of the Raman modulation frequency $\omega_m$. 
We assume that the Raman sidebands $0$ and $\nu$ are initially  
in independent quantum states $\hat\rho_{\mathrm{in}}^{(0)}$ and $\hat\rho_{\mathrm{in}}^{(\nu)}$, respectively, while the other sidebands are initially in the vacuum state. 
The density matrix of the initial  state of the  fields is given by 
\begin{equation}
\hat\rho_{\mathrm{in}}=\hat\rho_{\mathrm{in}}^{(0)}\otimes\hat\rho_{\mathrm{in}}^{(\nu)}\otimes\prod_{q\not=0,\nu} (|0\rangle\langle 0|)_q.
\label{18}
\end{equation}
Here, $\nu\not=0$.

\subsection{Modulation of photon statistics}

We study the mixing and modulation of photon statistics of the sideband fields.
When we use Eq.~(\ref{8})  to calculate the mean photon numbers of the sidebands generated from the initial
state (\ref{18}), we find 
\begin{eqnarray}
\langle \hat b_q^\dagger\hat b_q\rangle&=&
J_q^2(\kappa L)\langle\hat b_0^\dagger(0)\hat b_0(0)\rangle
+J_{q-\nu}^2(\kappa L)\langle\hat b_\nu^\dagger(0)\hat b_\nu(0)\rangle
\nonumber\\&&\mbox{}
+J_q(\kappa L)J_{q-\nu}(\kappa L)
\nonumber\\&&\mbox{}\times
[e^{-i\nu\pi/2}
\langle\hat b_0^\dagger(0)\rangle\langle\hat b_\nu(0)\rangle
+\mathrm{c.c.}] .
\label{19}
\end{eqnarray} 
Furthermore, we find
\begin{eqnarray}
\lefteqn{\langle \hat b_q^{\dagger 2}\hat b_q^2\rangle=
J_q^4(\kappa L)\langle\hat b_0^{\dagger 2}(0)\hat b_0^2(0)\rangle
+J_{q-\nu}^4(\kappa L)\langle\hat b_\nu^{\dagger 2}(0)\hat b_\nu^2(0)\rangle}
\nonumber\\&&\mbox{}
+4J_q^2(\kappa L)J_{q-\nu}^2(\kappa L)\langle\hat b_0^{\dagger}(0)\hat b_0(0)\rangle\langle\hat b_\nu^{\dagger}(0)\hat b_\nu(0)\rangle
\nonumber\\&&\mbox{}
+[2e^{-i\nu\pi/2}J_q^3(\kappa L)J_{q-\nu}(\kappa L)\langle\hat b_0^{\dagger 2}(0)\hat b_0(0)\rangle\langle\hat b_\nu(0)\rangle
\nonumber\\&&\mbox{}
+2e^{-i\nu\pi/2}J_q(\kappa L)J_{q-\nu}^3(\kappa L)\langle\hat b_0^\dagger(0)\rangle
\langle\hat b_\nu^\dagger(0)\hat b_\nu^2(0)\rangle
\nonumber\\&&\mbox{}
+e^{-i\nu\pi}J_q^2(\kappa L)J_{q-\nu}^2(\kappa L)\langle\hat b_0^{\dagger 2}(0)\rangle\langle\hat b_\nu^2(0)\rangle
+\mathrm{c.c.}]. \qquad
\label{20b}
\end{eqnarray}
Hence, the second-order autocorrelation function 
$\Gamma_q^{(2)}=\langle \hat b_q^{\dagger 2}\hat b_q^2\rangle-\langle \hat b_q^\dagger\hat b_q\rangle^2$ of the sideband $q$ is found to be  
\begin{eqnarray}
\Gamma_q^{(2)}&=&
J_q^4(\kappa L)\Gamma_0^{(2)}(0)
+J_{q-\nu}^4(\kappa L)\Gamma_\nu^{(2)}(0)
\nonumber\\&&\mbox{}
+2J_q^2(\kappa L)J_{q-\nu}^2(\kappa L)\Delta_0
\nonumber\\&&\mbox{}
+[e^{-i\nu\pi}J_q^2(\kappa L)J_{q-\nu}^2(\kappa L)\Delta_1
\nonumber\\&&\mbox{}
+2e^{-i\nu\pi/2}J_q^3(\kappa L)J_{q-\nu}(\kappa L)\Delta_2
\nonumber\\&&\mbox{}
+2e^{-i\nu\pi/2}J_q(\kappa L)J_{q-\nu}^3(\kappa L)\Delta_3
+\mathrm{c.c.}],
\nonumber\\
\label{20a}
\end{eqnarray}
where
\begin{eqnarray}
\Delta_0&=&\langle\hat b_0^{\dagger}(0)\hat b_0(0)\rangle\langle\hat b_\nu^{\dagger}(0)\hat b_\nu(0)\rangle-|\langle\hat b_0(0)\rangle|^2|\langle\hat b_\nu(0)\rangle|^2,
\nonumber\\
\Delta_1&=&\langle\hat b_0^{\dagger 2}(0)\rangle\langle\hat b_\nu^2(0)\rangle
-\langle\hat b_0^\dagger(0)\rangle^2\langle\hat b_\nu(0)\rangle^2,
\nonumber\\
\Delta_2&=&\langle\hat b_\nu(0)\rangle[\langle\hat b_0^{\dagger 2}(0)\hat b_0(0)\rangle
-\langle\hat b_0^\dagger(0)\hat b_0(0)\rangle\langle\hat b_0^\dagger(0)\rangle],
\nonumber\\
\Delta_3&=&\langle\hat b_0^\dagger(0)\rangle[
\langle\hat b_\nu^\dagger(0)\hat b_\nu^2(0)\rangle
-\langle\hat b_\nu^\dagger(0)\hat b_\nu(0)\rangle\langle \hat b_\nu(0)\rangle].
\end{eqnarray}
The first two terms on the right-hand sides of Eqs.~(\ref{19}), (\ref{20b}), and (\ref{20a}) are the contributions of the individual input sidebands $0$ and $\nu$. 
The other terms result from the interference between the two interaction channels. 

\begin{figure}
\begin{center}
\includegraphics{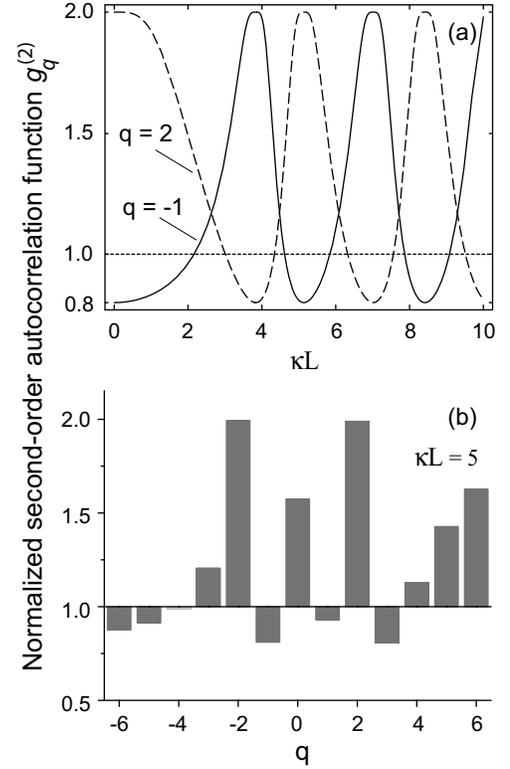}
\end{center}
\caption{
Normalized second-order autocorrelation function $g_q^{(2)}$ 
as a function of (a) the effective medium length $\kappa L$  and (b) the sideband order $q$ in the case
of two-mode input.
The input sideband $0$ is initially prepared in a Fock state, with $5$ photons.
The input sideband $1$ is initially prepared in a thermal state, with 1 photon in average. 
In (a), the sideband order is $-1$ (solid line) and $2$ (dashed line).
In (b), the effective medium length is $\kappa L =5$.} 
\label{Fig4}
\end{figure}

Unlike the case of single-mode input, in the case of two-mode input,
the normalized second-order autocorrelation function 
$g_q^{(2)}=1+\Gamma_q^{(2)}/\langle \hat n_q\rangle^2$    depends, in general, on $\kappa L$ and $q$.
Such behavior is illustrated in Fig.~\ref{Fig4}.
When $\kappa L$ is such that $J_q(\kappa L)=0$ or $J_{q-\nu}(\kappa L)=0$, we have 
$g_q^{(2)}=g_\nu^{(2)}(0)$ or $g_q^{(2)}=g_0^{(2)}(0)$, respectively. 
Consequently, if the two input sidebands have different normalized autocorrelation functions, i.e., $g_0^{(2)}(0)\not=g_\nu^{(2)}(0)$, then, with increasing $\kappa L$ or $q$,
the normalized autocorrelation function $g_q^{(2)}$ will oscillate
between the values $g_0^{(2)}(0)$ and $g_\nu^{(2)}(0)$ [see Fig.~\ref{Fig4}]. 
In particular, if the photon statistics of one of the input fields, e.g., the sideband $0$, 
is sub-Poissonian [$g_0^{(2)}(0)<1$]  and that of the other input field
is super-Poissonian [$g_\nu^{(2)}(0)>1$], then 
each generated  sideband $q$ will have complex statistical properties and will oscillate between  sub-Poissonian [$g_q^{(2)}<1$]
and super-Poissonian [$g_q^{(2)}>1$] photon statistics [see Fig.~\ref{Fig4}]. 
Using the prepared Raman coherence $\rho_0$ or the medium length $L$ as a control parameter, we can switch a sideband field from super-Poissonian photon statistics to sub-Poissonian or vice versa. 
Similar modulation of photon statistics has been demonstrated 
in a linear directional coupler \cite{coupler}. 

\subsection{Modulation of squeezing}

We study the mixing and modulation of the squeezing properties of the sideband fields.
When we use Eq.~(\ref{8})  to calculate the amplitudes 
$\langle \hat b_q\rangle$ and $\langle \hat b_q^2\rangle$ of the sidebands generated from the initial
state (\ref{18}), we find the  expressions
\begin{equation}
\langle\hat b_q\rangle=e^{iq\pi/2}J_q(\kappa L)\langle\hat b_0(0)\rangle
+e^{i(q-\nu)\pi/2}J_{q-\nu}(\kappa L)\langle\hat b_\nu(0)\rangle
\label{20}
\end{equation}
and 
\begin{eqnarray}
\langle\hat b_q^2\rangle&=&e^{iq\pi}J_q^2(\kappa L)\langle\hat b_0^2(0)\rangle
+e^{i(q-\nu)\pi}J_{q-\nu}^2(\kappa L)\langle\hat b_\nu^2(0)\rangle
\nonumber\\&&\mbox{}
+2e^{i(q-\nu/2)\pi}J_q(\kappa L)J_{q-\nu}(\kappa L)
\langle\hat b_0(0)\rangle\langle\hat b_\nu(0)\rangle.
\nonumber\\
\label{21}
\end{eqnarray}
We insert Eqs.~(\ref{19}), (\ref{20}), and (\ref{21}) into Eq.~(\ref{12c}). Then, we obtain  the squeezing
factor
\begin{eqnarray}
S_q(\varphi+q\pi/2)&=&J_q^2(\kappa L) S_0^{\mathrm{(in)}}(\varphi)
\nonumber\\&&\mbox{}
+J_{q-\nu}^2(\kappa L) S_\nu^{\mathrm{(in)}}(\varphi+\nu\pi/2),
\label{22}
\end{eqnarray}
where $S_0^{\mathrm{(in)}}$ and $S_\nu^{\mathrm{(in)}}$ are the initial squeezing factors of the  sidebands $0$ and $\nu$, respectively. As seen, the squeezing factor $S_q$ of the sideband $q$ is  a superposition of the input squeezing factors $S_0^{\mathrm{(in)}}$ and $S_\nu^{\mathrm{(in)}}$, taken
with the  quadrature phase shifts $-q\pi/2$ and $-(q-\nu)\pi/2$, respectively,
and weighted by the factors $J_q^2(\kappa L)$ and 
$J_{q-\nu}^2(\kappa L)$, respectively. 

\begin{figure}
\begin{center}
\includegraphics{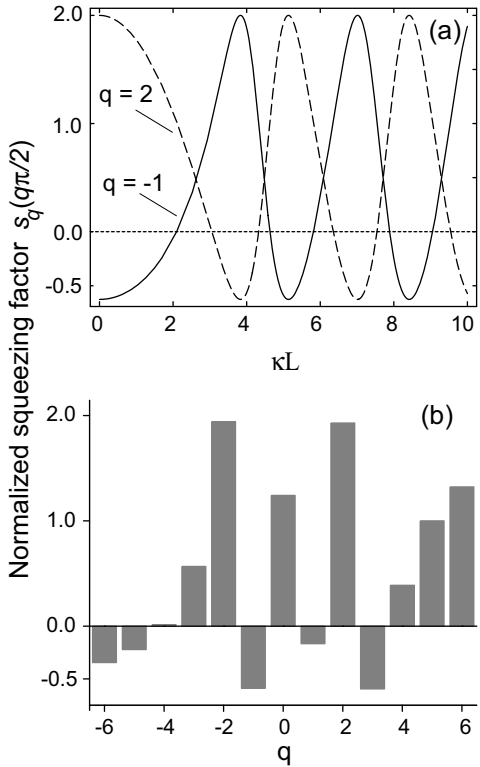}
\end{center}
\caption{
Normalized squeezing factor $s_q(q\pi/2)$ 
as a function of (a) the effective medium length $\kappa L$  and (b) the sideband order $q$ in the case
of two-mode input.
The input sideband $0$ is initially prepared in a squeezed vacuum state, with the squeezing parameter $r=1$
and the phase $\theta=0$.
The input sideband $1$ is initially prepared in a thermal state, with the mean photon number 1. 
In (a), the sideband order is $-1$ (solid line) and $2$ (dashed line).
In (b), the effective medium length is $\kappa L =5$.} 
\label{Fig5}
\end{figure}

Unlike the case of single-mode input, in the case of two-mode input, 
the normalized squeezing factor 
$s_q(\varphi+q\pi/2)=S_q(\varphi+q\pi/2)/\langle\hat{n}_q\rangle$    varies, in general, with $\kappa L$ and $q$.
Such behavior is illustrated in Fig.~\ref{Fig5}.
When $\kappa L$ is such that $J_q(\kappa L)=0$ or $J_{q-\nu}(\kappa L)=0$, we have 
$s_q(\varphi+q\pi/2)=s_\nu^{\mathrm{(in)}}(\varphi+\nu\pi/2)$ 
or $s_q(\varphi+q\pi/2)=s_0^{\mathrm{(in)}}(\varphi)$, respectively.
Consequently, if the normalized squeezing factors $s_0^{\mathrm{(in)}}(\varphi)$ and $s_\nu^{\mathrm{(in)}}(\varphi+\nu\pi/2)$ of the two input fields are different, 
the normalized squeezing factor $s_q(\varphi+q\pi/2)$ will oscillate
between the values $s_0^{\mathrm{(in)}}(\varphi)$ and $s_\nu^{\mathrm{(in)}}(\varphi+\nu\pi/2)$. 
In particular, if one of the two input fields, e.g., the sideband $0$, is squeezed [$s_0^{\mathrm{(in)}}(\varphi_0)<0$] and the other input field is not squeezed [$s_\nu^{\mathrm{(in)}}(\varphi)>0$], then, each generated  sideband $q$ will have complex 
squeezing properties and will oscillate between a squeezed state 
[$s_q(\varphi_0+q\pi/2)<0$] and a non-squeezed state [$s_q(\varphi)>0$],   see Fig.~\ref{Fig5}. 
Using the prepared Raman coherence $\rho_0$ or the medium length $L$ as a control parameter, 
we can switch a sideband field from a non-squeezed state 
to a squeezed state  or vice versa. 
Note that a similar result has been obtained for a linear directional
coupler \cite{coupler}.

\begin{figure}
\begin{center}
\includegraphics{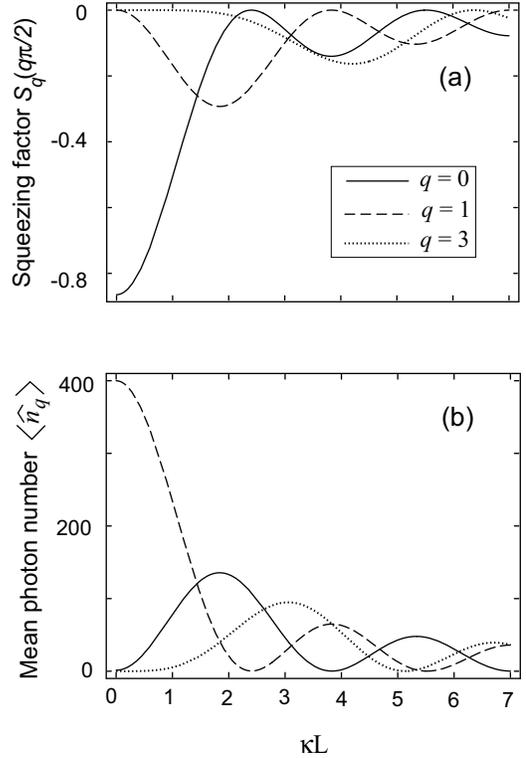}
\end{center}
\caption{
(a) Squeezing factor $S_q(q\pi/2)$ and (b) mean photon number $\langle\hat n_q\rangle$ 
as  functions of the effective medium length $\kappa L$ in the case
where the sideband $0$ is initially prepared in a weak squeezed vacuum state and the sideband  
$1$ is initially prepared in a strong coherent state. 
The parameters for the initial states of the input sidebands are $r=1$, $\theta=0$, and $\alpha= 20$.
The curves are calculated for the sidebands $0$ (solid line), $1$ (dashed line), and $3$ (dotted line).} 
\label{Fig6}
\end{figure}

We  analyze a particular case where the sideband $0$ is initially in a squeezed vacuum state $|\xi\rangle_0=
\exp[(\xi^*\hat b_0^2-\xi\hat b_0^{\dagger 2})/2]|0\rangle_0$ and  
the sideband $\nu$ is initially in a coherent state $|\alpha\rangle_\nu$. 
Here, $\xi=re^{i\theta}$ is a complex number, the modulus $r=|\xi|$ characterizes the amount
of squeezing, and the phase angle $\theta$ characterizes the alignment of the squeezed vacuum state in phase space.
For the input  sideband $0$, 
we have \cite{Mandel and Scully book}
$\langle\hat b_0^\dagger(0)\hat b_0(0)\rangle=\sinh^2 r$, $\langle\hat b_0(0)\rangle=0$, and 
$S_0^{\mathrm{(in)}}(\varphi)=\cosh 2r -\cos(2\varphi-\theta)\sinh 2r-1 $.  
For the input sideband $\nu$, we have 
$\langle\hat b_\nu^\dagger(0)\hat b_\nu(0)\rangle=|\alpha|^2$, $\langle\hat b_\nu(0)\rangle=\alpha$, and $S_\nu^{\mathrm{(in)}}(\varphi)=0$.
Then, we find
from Eq.~(\ref{19}) that the mean photon number of an arbitrary sideband $q$ is 
\begin{equation}
\langle \hat b_q^\dagger\hat b_q\rangle=J_q^2(\kappa L) \sinh^2 r +J_{q-\nu}^2(\kappa L)|\alpha|^2.
\label{23}
\end{equation}
We find from Eq.~(\ref{22}) that the maximal squeezing of the sideband $q$ occurs in the $\varphi_q$-quadrature where $\varphi_q=\theta/2+q\pi/2$. 
The corresponding value of the squeezing factor is  
\begin{equation}
S_q(\varphi_q)=J_q^2(\kappa L)(e^{-2r}-1).
\label{24}
\end{equation}
As seen from Eq.~(\ref{24}), squeezing can be transferred from the initial squeezed vacuum state of the sideband
$0$ to the other sidebands. 
The squeezing factors of the sidebands  are independent of the amplitude $\alpha$ of the initial coherent state of the sideband $\nu$.
Meanwhile, the mean photon number of each sideband is governed not only by the squeezing parameter $r$ of
the initial state of the sideband $0$ but also by the amplitude $\alpha$ of the initial state of the sideband $\nu$.
Using this fact, we can manipulate to get optimized mean photon numbers and 
squeezing degrees of the sideband fields at the output as per requirement.
In particular, we can convert squeezing from a weak field to a much stronger field.
To illustrate this possibility, we plot in Fig.~\ref{Fig6} the  squeezing factor $S_q(q\pi/2)$ and 
the mean photon number $\langle\hat n_q\rangle$ as  functions of the effective medium length $\kappa L$
for the parameters   $r=1$, $\theta=0$, $\nu=1$, and $\alpha=20$. 
In this case, the most negative value of the input squeezing factor $S_0^{\mathrm{(in)}}(\varphi)$ is achieved at $\varphi=0$ and is given by
$S_0^{\mathrm{(in)}}(0)=e^{-2r}-1\cong -0.86$, indicating the squeezing degree  86\%.
The mean photon number of the input squeezed vacuum state is 
$\langle\hat b_0^\dagger(0)\hat b_0(0)\rangle=\sinh^2 r\cong 1.38$, rather small.
The solid lines in Fig.~\ref{Fig6} show that the sideband $0$, initially prepared in a weak squeezed
vacuum state,
can be significantly enhanced while keeping its squeezing degree substantial. 
Meanwhile, the dashed lines  show that, for $\kappa L=1.84$, 
the sideband $1$, initially prepared in a strong coherent state, is squeezed by about  29\% 
and has the mean photon number of about 41.  
Similarly, the dotted lines  show that, for $\kappa L=4.2$, a  generated new sideband $3$
is squeezed by about 16\% and has the mean photon number of about 39.  
Thus, from a weak squeezed field at the input,  we can obtain   
other output squeezed fields that have  smaller but still  substantial squeezing degrees, 
much larger mean photon numbers, and  different frequencies. 

\subsection{Two-photon interference}

We show the possibility of quantum interference between the probability amplitudes for
a pair of photons with different frequencies in
the coherent Raman process.
We assume that the sidebands $0$ and $1$ are initially prepared in  independent single-photon states. This initial
condition corresponds to  the situation where 
two photons  with different frequencies $\omega_0$ and  $\omega_1$ are incident into the Raman medium. 
The input state of the fields can be written as
\begin{equation}
|\Psi_{\mathrm{in}}\rangle=|1\rangle_0|1\rangle_1\prod_{q\not=0,1}|0\rangle_q =
\hat b_0^\dagger(0)\hat b_1^\dagger(0) |0\rangle.
\end{equation}
The output state of the fields is given by $|\Psi_{\mathrm{out}}\rangle=
\hat b_0^\dagger(-L/c)\hat b_1^\dagger(-L/c) |0\rangle$. With the help of Eq.~(\ref{8}), we find
\begin{eqnarray}
|\Psi_{\mathrm{out}}\rangle&=&-i\sqrt2\sum_q e^{iq\pi} J_q(\kappa L) J_{q-1}(\kappa L) |2_q\rangle
\nonumber\\&&\mbox{}
-i\sum_{k<l}e^{i(k+l)\pi/2}[J_k(\kappa L) J_{l-1}(\kappa L)
\nonumber\\&&\mbox{}
 + J_l(\kappa L) J_{k-1}(\kappa L)]|1_k 1_l\rangle.
\label{16a}
\end{eqnarray}
Here, the Fock state $|2_q\rangle$ is the state of two photons in the sideband $q$ with no photons in the other sidebands,
and the Fock state $|1_k 1_l\rangle$ is the state in which there is one photon in each of the  sidebands $k$ and $l$ but no photons in the other sidebands. 

It follows from Eq.~(\ref{16a}) that the probability for finding
two photons in the sideband $q$ is
\begin{equation}
W_q^{(2)}=2J_q^2(\kappa L) J_{q-1}^2(\kappa L).
\end{equation}
The joint probability  for finding one photon in each of the sidebands $k$ and $l$ ($k\not=l$) is given by  
\begin{equation}
W_{kl}=[J_k(\kappa L) J_{l-1}(\kappa L)
+ J_l(\kappa L) J_{k-1}(\kappa L)]^2.
\label{16d}
\end{equation}
The probability $W_q^{(1)}=\sum_{l\not=q}W_{ql}$  for having one and only one photon in the sideband $q$ is 
\begin{equation}
W_q^{(1)}=J_q^2(\kappa L)+J_{q-1}^2(\kappa L)
-4J_q^2(\kappa L)J_{q-1}^2(\kappa L).
\end{equation}
The mean photon number of the sideband $q$ is 
\begin{equation}
\langle\hat n_q\rangle=J_q^2(\kappa L)+J_{q-1}^2(\kappa L).
\end{equation}
We find the relations $W_{-q}^{(2)}=W_{1+q}^{(2)}$, $W_{-q}^{(1)}=W_{1+q}^{(1)}$, and
$\langle\hat n_{-q}\rangle=\langle\hat n_{1+q}\rangle$,
which reflect the symmetry of the
generated Stokes and anti-Stokes sidebands with respect to the two input sidebands 0 and 1.

\begin{figure}
\begin{center}
\includegraphics{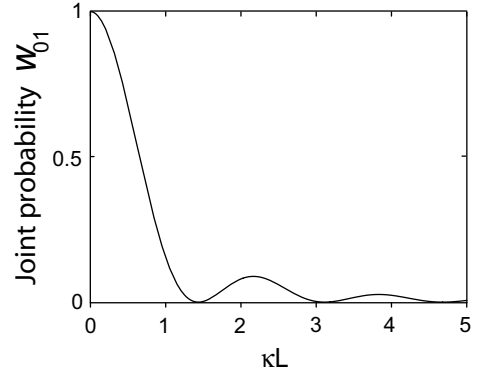}
\end{center}
\caption{Joint probability  $W_{01}$ for finding one photon in each of the sidebands $0$ and $1$ as 
a function of the effective medium length $\kappa L$.} 
\label{Fig7}
\end{figure}

When we insert $k=0$ and $l=1$ into Eq.~(\ref{16d}), we obtain the following expression
for the joint probability  for finding one photon in each of the sidebands $0$ and $1$:
\begin{eqnarray}
W_{01}&=&[J_0^2(\kappa L)- J_1^2(\kappa L)]^2.
\end{eqnarray}
This expression shows that  
the joint probability $W_{01}$  may become zero at certain values of $\kappa L$
[see Fig.~\ref{Fig7}].
This is a signature of  destructive interference between two channels that form the state $|1_01_1\rangle$.
In the first channel, each of the two photons individually  transmits through the medium without any changes.  
The two-photon probability amplitude for this channel is 
$J_0(\kappa L)J_0(\kappa L)=J_0^2(\kappa L)$.
In the second channel, both the photons are scattered from the prepared Raman coherence and
exchange their sidebands.
The two-photon probability amplitude for this channel is 
$e^{i\pi/2}J_1(\kappa L)e^{i\pi/2}J_{1}(\kappa L)=- J_1^2(\kappa L)$. 
Since Raman scattering produces a  phase shift of $\pi/2$ for each photon, 
the probability amplitudes for the two channels (the transmission and scattering of both the photons) are $180^\circ$  out of phase. 
The interference between the two channels is therefore destructive, yielding the output state $|1_01_1\rangle$ 
with the joint probability  $W_{01}$ given above. 
When the  medium length $L$ is such that $J_0^2(\kappa L)=J_1^2(\kappa L)$, the interference between the two two-photon amplitudes  becomes completely destructive, and therefore
the state $|1_01_1\rangle$ is removed from the output state (\ref{16a}). 
We denote such a medium length by $L_f$. 
The positions of the zeros of  $W_{01}$ depicted in Fig.~\ref{Fig7} indicate that the first three values of $L_f$ are given by $\kappa L_f=1.44$, 3.11, and 4.68.
It is interesting to note that $\kappa L_f$ can be determined in an experiment
using a single-mode input. Indeed, in the case where a single sideband $0$ is initially excited,  
the mean photon numbers
of the generated sidebands are given by Eq.~(\ref{9a}). Therefore, the effective medium length $\kappa L_f$ corresponds to the
situation where  the probe sideband $0$ and its adjacent sidebands $\pm1$
have the same mean photon numbers at the output.

There exist literatures on two-photon interference in various systems \cite{Hong,applications,Qbeat}.
Two-photon interference in coherent Raman scattering, described above, is an analogy of two-photon interference at a conventional beam splitter \cite{Hong,Mandel and Scully book}. 
We emphasize that two-photon interference in coherent Raman scattering involves copropagating photons with different frequencies in a collinear scheme.

\subsection{General relation between the $P$ representations of the input and output states}

To be more general,  
we consider the case where an arbitrary number of  sidebands is initially excited. 
We find that an arbitrary multimode coherent state 
$|\{\alpha_q(0)\}\rangle$ of the input fields produces a  coherent state 
$|\{\alpha_q(L/c)\}\rangle$ of the output fields. Here, the output amplitudes $\{\alpha_q(L/c)\}$
are linearly transformed from the input amplitudes $\{\alpha_q(0)\}$ as given by 
\begin{equation}
\alpha_q(L/c)=\sum_{q'}e^{i(q-q')\pi/2}J_{q-q'}(\kappa L) \alpha_{q'}(0). 
\label{16}
\end{equation}
Consequently, the diagonal coherent-state representation $P_{\mathrm{in}}(\{\alpha_q\})$ of an arbitrary input quantum state
$\hat\rho_{\mathrm{in}}$  determines  the  representation $P_{\mathrm{out}}(\{\alpha_q\})$ 
of the output state $\hat\rho_{\mathrm{out}}$    via the equation 
\begin{equation}
P_{\mathrm{out}}(\{\alpha_q\})=P_{\mathrm{in}}(\{\alpha'_q\}).
\label{17}
\end{equation}
Here, we have introduced the notation
\begin{equation}
\alpha'_q=\sum_{q'}e^{-i(q-q')\pi/2}J_{q-q'}(\kappa L)\alpha_{q'}.
\label{17a}
\end{equation}

If the input state $\hat\rho_{\mathrm{in}}$ is a classical state \cite{Mandel and Scully book}, $P_{\mathrm{in}}(\{\alpha_q\})$ must be 
non-negative  and less singular than a $\delta$ function,  and 
consequently so must $P_{\mathrm{out}}(\{\alpha_q\})$.
In this case, the output state $\hat\rho_{\mathrm{out}}$ is also a classical state. Moreover,  
since the multimode coherent state $|\{\alpha_q\}\rangle$ is separable and the weight factor $P_{\mathrm{out}}(\{\alpha_q\})$ is non-negative, the output
state $\hat\rho_{\mathrm{out}}$ is,  by definition, separable \cite{Chuang,Zeilinger}. 
Therefore, a necessary condition for the output fields
to be in an inseparable (entangled) state or, more generally, in a nonclassical state 
is that the input field state is  a nonclassical state. 
A similar condition has been derived for the beam splitter entangler \cite{Knight}. 
Note that, in the case where we use a
single-mode input field $q=0$, prepared in an arbitrary  quantum state with the coherent-state representation $P_{\mathrm{in}}^{(0)}(\alpha)$  
(the Stokes and anti-Stokes  sideband fields are initially in the vacuum state), 
Eq.~(\ref{17}) becomes
\begin{equation}
P_{\mathrm{out}}(\{\alpha_q\})=
P_{\mathrm{in}}^{(0)}(\alpha'_0)
\prod_{q\not=0}
\delta(\alpha'_q).
\end{equation}
It has been shown in Ref.~\cite{entang} that, when the input field is prepared in an even or odd coherent state,
a multipartite entangled coherent state can be generated.

\section{Conclusions and discussions}
\label{sec:summary}

We have studied the quantum properties of  multiorder sidebands generated by the beating  
of a quantum probe field with a prepared Raman coherence in a far-off-resonance medium. 
Under the conditions of negligible dispersion and limited
bandwidth, we have derived a  Bessel-function solution for the sideband field operators.
We have analytically and numerically calculated various quantum statistical characteristics of
the multiorder sideband fields. 

We have examined the quantum properties of the sideband fields in the case of  single-mode quantum input. 
We have shown that,
when we change the effective medium length or the Raman sideband order, the 
autocorrelation functions, the cross-correlation functions, 
the photon-number distributions, and the squeezing factors undergo oscillations governed by the Bessel functions. 
When the  sideband order is higher,  the  onset of the sideband generation occurs later 
and, therefore, so does  the onset of the sideband autocorrelation functions.
The mean photon number  and the autocorrelation functions 
of each sideband reach their largest values at the same optimal medium length determined by  
the first peak of the corresponding Bessel function. 
The higher the sideband order, the larger is the optimal length and the smaller is the  maximal output values of the mean photon number and sideband autocorrelation functions.

Meanwhile, the normalized autocorrelation functions and normalized squeezing factors  
of the probe field  
are not altered by the parametric beating process.  They are replicated 
into the comb of generated  multiorder sidebands. 
As the result, the single-mode input field and the generated sidebands 
have identical normalized autocorrelation functions and identical normalized squeezing factors.
In other words, they have similar quantum statistical properties -- the same type of photon
statistics and the same type of squeezing.
In addition to this resemblance, 
it has been shown that an input field in a coherent state can produce sideband fields in coherent states.
It has also been shown that, when the input field is prepared in a thermal state, the reduced state of each
generated sideband is also a thermal state.
Therefore, the multiorder coherent Raman
process can be used to  multiplex the statistical properties of a quantum probe field into a broad comb of 
different frequencies. 

As far as replicating the statistical properties of the input probe into its sidebands is concerned,
the Raman medium appears to behave as a linear system.
However, the replication of the normalized autocorrelation functions  and normalized squeezing factors of the 
probe field does not mean the replication of the quantum state at all. 
The photon-number distributions and the quantum states of the sidebands evolve in a rather complicated way.
Cross-correlations between the sidebands can be generated from  initially uncorrelated fields. 
An inseparable state can be generated from a separable nonclassical state.
Although the dynamics of our model system is linear with respect to the field variables, the possibilities of interesting quantum phenomena such as anti-correlation generation,  squeezing multiplexing, and
entangled-state generation represent the quantum properties that 
cannot be described by the classical statistics of the fields with positive $P$ functions. 

We have also studied the mixing and modulation of photon statistical properties in the case of
two-mode quantum input. We have shown that   
the prepared Raman coherence and  the medium length can be used as control parameters to switch a sideband field from one type of photon statistics to another type, or from a non-squeezed state to a squeezed
state and vice versa.  In addition, we can switch  nonclassical properties, such as sub-Poissonian photon statistics and squeezing,
from one frequency to another frequency.
We have also shown an example of quantum interference between the probability amplitudes for
a pair of photons with different frequencies.
 
We have made interesting observations that 
the multiorder coherent Raman scattering behaves in many aspects as a conventional beam splitter and hence
can be called a multiorder Raman beam splitter.
The two systems have the same underlying physics:  the fields are linearly transformed
from the input values. However, the two systems are different in their natures.
Unlike the conventional beam splitter, the multiorder coherent Raman process can efficiently produce
a broad comb of sideband fields whose frequencies are different and are separated by integer multiples of
the Raman modulation frequency. 
The number of generated Raman sidebands increases with the effective medium length, which
is  proportional to the product of the medium length and the prepared Raman coherence. 
The Bessel functions of
the effective medium length  play 
a similar role as the transmission and reflection coefficients of a conventional beam splitter. 

The ability of the far-off-resonance Raman medium to generate a broad comb of fields with similar quantum statistical properties and
to switch the quantum statistical characteristics of the radiation fields from one type to another type
may find useful applications for high-performance optical communication networks. 
In addition,
two-photon interference in coherent Raman scattering may find
various  applications for high-precision measurements and also for quantum computation.

Finally, we emphasize that the  coupling between the Raman sidebands can be  controlled 
by the magnitude of the
prepared Raman coherence, that is, by the intensities of the driving fields. 
In a realistic far-off-resonance Raman medium, such as
molecular hydrogen or deuterium vapor \cite{Modulation,D2}, solid hydrogen \cite{subfem,Kien99}, and rare-earth doped dielectrics \cite{kolesov},
a large Raman coherence  and, consequently, 
a large number of Raman sidebands can be generated by the two-color adiabatic pumping technique. 
In such a system, the generation of a broad comb of high-order Raman sidebands with nonclassical statistical properties is, in principle, feasible. 
Therefore, we expect that the coherent Raman scattering technique using quantum fields will 
become a practical and efficient method for a wide
range of applications in nonlinear and quantum optics.

\end{document}